\documentclass[aps,prl,amsmath,amssymb,twocolumn,superscriptaddress,showpacs]{revtex4-1}
\usepackage{graphicx}
\usepackage{hyperref}
\hypersetup{
	colorlinks=true,       
	linkcolor=blue,          
	citecolor=blue,        
	filecolor=magenta,      
	urlcolor=blue         
}
\usepackage{natbib}

\begin{document}

\title{Quantum Enhanced Magnetometer with Low-Frequency Squeezing}

\author{Travis Horrom}
\affiliation{Department of Physics, The College of William and Mary, Williamsburg, VA 23187, USA}

\author{Robinjeet Singh}
\affiliation{Hearne Institute of Theoretical Physics and Department of Physics and Astronomy, Louisiana State University, Baton Rouge, LA 70803, USA}

\author{Jonathan P. Dowling}
\affiliation{Hearne Institute of Theoretical Physics and Department of Physics and Astronomy, Louisiana State University, Baton Rouge, LA 70803, USA}
\affiliation{Beijing Computational Science Research Center, Beijing, 100084, China}

\author{Eugeniy E. Mikhailov}
\affiliation{Department of Physics, The College of William and Mary, Williamsburg, VA 23187, USA}

\date{\today}

\begin{abstract}

 We report the demonstration of a magnetometer with noise-floor reduction 
 below the shot-noise level.  This magnetometer, based on a nonlinear
 magneto-optical rotation effect, is enhanced by the injection of
 a squeezed vacuum state into its input. The noise spectrum shows squeezed
 noise reduction
 of about 2$\pm 0.35$~dB spanning from close to 100 Hz to several megahertz.  We also 
 report 
 on the observation of two different regimes of operation of  such a
 magnetometer: one in which the
 detection noise is limited by the quantum noise of the light probe only,
 and one in which we see additional noise originating
 from laser noise which is rotated into the vacuum polarization.

\end{abstract}


\maketitle

Optical magnetometers now  reach
the sub-femtotesla/$\sqrt{\text{Hz}}$ level of
sensitivity~\cite{polzik2010prl,romalis2007natphys}, surpassing superconducting
quantum interference device (SQUID) magnetometers~\cite{seton2005cryogenic}.
Ultimately, such optical magnetometers are limited by quantum-mechanical noise
sources, in particular by the photon shot-noise at detection,  spin
projection atomic noise, and the back action of light noise 
onto atoms~\cite{fleischhauer2000pra_ac_shtark_shifts, polzik2010prl, romalis2007natphys}. 
The former noise source can be addressed with
injection of polarization-squeezed light states~\cite{mitchel2010prl_sqz}, while
the spin projection noise can be suppressed via the use of atoms prepared in
spin-squeezed states~\cite{kitagawa1993pra_squeezed_spin, romalis2007natphys}
 or with quantum
nondemolition measurements~\cite{mitchel2010prl_sub_projection_noise,mitchel2011nature}.

In this manuscript, we demonstrate a quantum-enhanced, all-atomic optical
magnetometer based on a nonlinear magneto-optical (Faraday) rotation
(NMOR)~\cite{budker'98,budker'00,budker2002rmp,NovikovaW02,novikova_ac-stark_2000,novikova05josab},
with the injection of a vacuum-squeezed state into the polarization
orthogonal to that of the probe field.
We also demonstrate the transition from a shot-noise-limited
magnetometer at lower atomic densities, to a region where the 
magnetometer is affected by
the interaction of the light noise with the atoms at higher atomic densities.
In contrast to a previously reported magnetometer, with squeezing
generated via parametric down conversion in a nonlinear
crystal~\cite{mitchel2010prl_sqz}, our setup uses an atomic squeezer based on
the polarization self-rotation (PSR)
effect~\cite{MatskoNWBKR02, ries_experimental_2003, mikhailov2008ol,
mikhailov2009jmo, grangier2010oe, lezama2011pra, mikhailov2012sq_pulses}.
Unlike its crystal counterpart, the PSR squeezer does not require a powerful
pump laser, but uses a pump laser with only several milliwatts of power 
in a single-path configuration. 
While the original simple model predicts about
6~dB of squeezing~\cite{MatskoNWBKR02} and a detailed treatment predicts
about 8~dB of squeezing with cold Rb atoms~\cite{mikhailov2011jmo}, the
best demonstrated squeezing via PSR in  hot Rb atoms so far is
3~dB~\cite{lezama2011pra}. 
Our
squeezer generates about 2~dB of noise reduction, starting from close to 100 Hz 
and ranging up to 
several megahertz. This is the lowest frequency quantum noise
sideband suppression generated at a wavelength of 795~nm to date. This unique 
squeezer allows for a
quantum enhanced all-atomic magnetometer with improvements to the
 signal-to-noise ratio for measurements in the 
same range of frequencies.  This is potentially useful for gravitational wave
detectors~\cite{harry2002prd},  geophysics, astronomy, biophysics,
and medical applications. It is particularly useful for detecting low-frequency magnetic signatures against a background of a constant field.



\begin{figure}[h]
	\begin{center}
		\includegraphics[width=1.0\columnwidth]{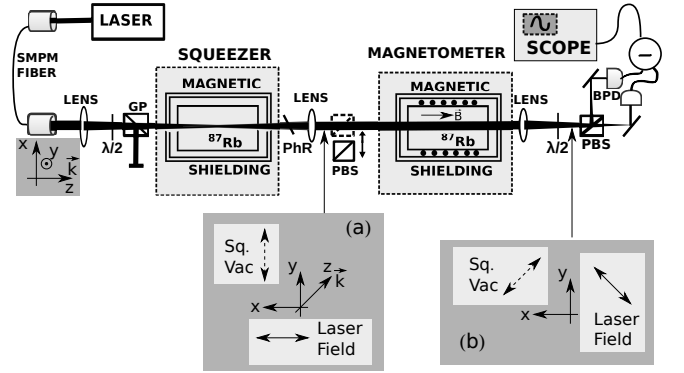}
	\end{center}
	\caption{
	Experimental setup. The squeezer prepares an optical field with reduced
	noise properties which is used as a probe for the magnetometer. SMPM
	fiber depicts single-mode polarization-maintaining fiber, $\lambda/2$ is
	half-wave plate, PhR is phase-retarding wave plate, PBS is polarizing beam
	splitter, GP is Glan-laser polarizer, BPD is balanced  photodetector.
 	Axes x and y coincide with horizontal and vertical
	polarization axes of all PBSs in our setup, axis z is along beam
	propagation direction.
	Inserts show the polarization of squeezed vacuum (Sq.Vac) field
	and laser field before the magnetometer cell (a) and right before the
	last PBS (b). 
	}
	\label{fig:setup}
\end{figure}

The setup of our experiment is depicted in Fig.~\ref{fig:setup}. It contains
two important components: the squeezer, which prepares the polarization-squeezed
probe beam, and the magnetometer, which can be probed with either the squeezed
or shot-noise-limited (unsqueezed) beam.

\begin{figure}[h]
	\begin{center}
		\includegraphics[width=1.0\columnwidth]{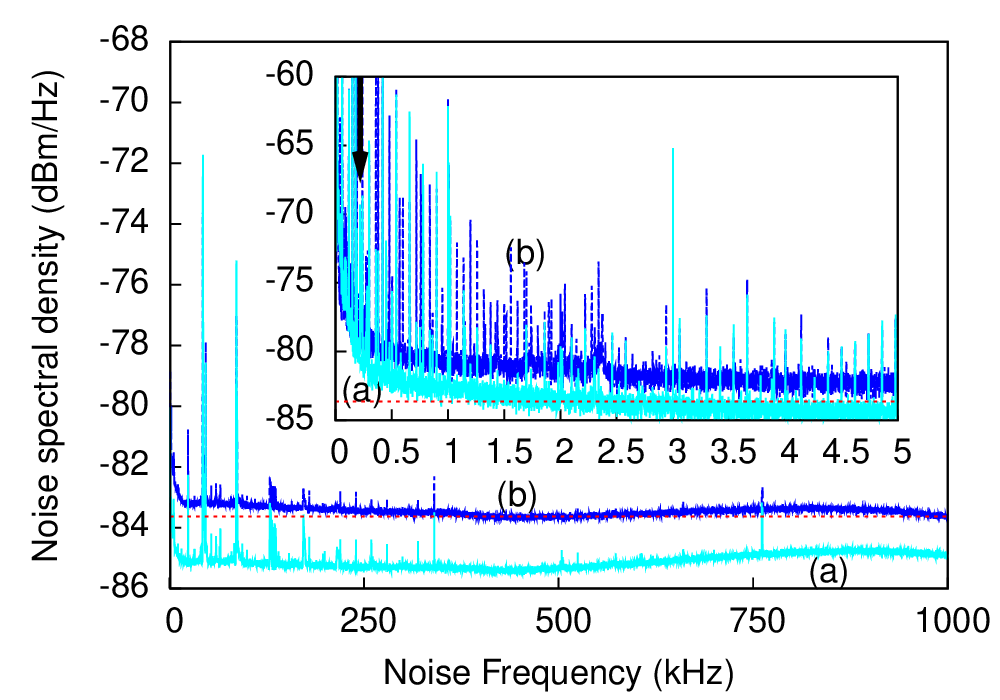}
	\end{center}
	\caption{
	(Color online)
	Magnetometer quantum noise spectrum with  polarization-squeezed(a)
	and  shot-noise-limited (b) probe fields taken at magnetometer cell
	temperature of 35$^\circ$C. The insert shows the low frequency part of the
	noise spectrum ( 0 to 5~kHz). The arrow
	marks the frequency of magnetic field modulation  at 220~Hz.
	Laser probe power is 6~mW.  Spectrum analyzer resolution bandwidth is 0.9~Hz.
	}
	\label{fig:lf_sqz}
\end{figure}

The operation of this squeezer is discussed in detail in Ref.~\cite{mikhailov2012sq_pulses}.  
The output of a DL100 
Toptica external  cavity semiconductor  laser, locked  to the D$_1$  line $F_g=2
\to  F_e=2$ transition  of $^{87}$Rb with zero detuning,  passes  through a 
single-mode
polarization-maintaining  (SMPM)  fiber  to  achieve  an  axially  symmetric
Gaussian  intensity distribution  of the  pump laser  beam. The  Glan-laser
polarizer (GP) purifies the polarization  of the pump beam and ensures its
linear x-polarization. The pump laser beam, with a power of 7~mW, is focused
inside 
the Rb cell (which
contains isotopically enriched $^{87}$Rb vapor  and no buffer gas) with a
beam waist of 100~$\mu$m.  The Pyrex cell has a length of
75~mm and is enclosed in three layers of  $\mu$-metal magnetic shielding to 
screen out
ambient laboratory  magnetic fields and guarantee zero field
inside the cell. We maintain  the cell at 66$^\circ$ Celsius, corresponding
to an atomic  number density of $5.4\times10^{11}$  atoms/cm$^3$. We find these
parameters 
experimentally to be optimal for noise suppression
(squeezing) of 2$\pm 0.35$~dB with respect to the shot-noise level at frequencies
in the
range of several kHz to 1~MHz, and once we account for detection noise, we
see noise suppression  to frequencies as low as 100~Hz (see Fig.~\ref{fig:lf_sqz}).
This squeezed vacuum state is linearly polarized in the y-direction
(orthogonal to the x-polarized pump laser field) as shown in Fig.~\ref{fig:setup}(a).
After the first
cell, we make a collimated magnetometer probe beam
from the squeezer output with a waist size of 900~$\mu$m. 
We must treat this probe
quantum mechanically and thus
describe quantum fluctuations in both x and y polarizations.
The mixing of the squeezed-vacuum  field in the y-polarization, with
the strong pump field in the orthogonal polarization, creates a polarization-squeezed state~\cite{parashchuk1993qe_polarization_squeezed_light}, as was
first demonstrated in~\cite{polzik2000jmo_sq_polarization}.
When we set a polarizing beam splitter at 45$^\circ$ with respect to
polarization of the squeezed vacuum (see Fig.~\ref{fig:setup}(b)),
and thus split the laser power 50/50 for the balanced photodetector (BPD),
we make the detector sensitive to the quantum fluctuations in the squeezed
vacuum field~\cite{polzik2000jmo_sq_polarization, lezama2011pra, mitchel2010prl_sqz}. 
We use  this polarization-squeezed
beam as  the probe  field for  our magnetometer and refer to it as the {\em
squeezed probe} everywhere in the text. The laser power of this squeezed
probe is 6~mW after absorption loss in the squeezing cell.

The magnetometer itself consists of a similar cell of isotopically enriched 
$^{87}$Rb with the addition of 2.5~Torr Ne buffer gas. This
cell is also enclosed in the magnetic shielding, but an internal solenoid
controls
the magnetic field ($\vec{B}$)  which is parallel to
the direction of probe beam propagation.
We also vary the magnetometer cell temperature to see what
density of atoms provides an optimal magnetometer response.

After the magnetometer cell, we have a detection scheme to measure 
the polarization rotation angle of the probe through the atoms.
The scheme consists of a polarizing beam splitter  (PBS) set to 45$^\circ$
with respect to  the probe light polarization, which splits  the probe field at a
50/50 ratio  and directs it  to the balanced  photodetector (BPD).  The signal
from the BPD is sent to an SRS SR560 voltage preamplifier 
and then to a Lecroy 640Zi
oscilloscope to analyze the response of the system to the magnetic field and
also
measure the quantum noise spectrum (with the spectrum analyzer feature enabled
by the scope).
We tilt the phase-retarding plate after the squeezer (implemented with a
quarter-wave plate set so that the axes
of  birefringence  coincide  with the polarizations of the probe and
squeezed fields) to control the 
phase shift between orthogonal polarizations and adjust the 
squeezing angle of the vacuum field relative to the probe  field. In this way,
we can choose
the phase-angle to achieve the
maximum quantum noise suppression.

We  can remove  the squeezed-vacuum field  from the squeezed-probe beam  by inserting a PBS
before 
the magnetometer,
which rejects squeezed vacuum in the y-polarization and thus creates a
shot-noise-limited, unsqueezed,
coherent vacuum quantum state in this polarization, orthogonal to the
x-polarized pump
laser. Meanwhile, it leaves the intensity and the quantum state along the
x-polarization of the probe
virtually unaffected (we disregard small optical losses inside the PBS). We use
this normal unsqueezed beam to calibrate the response of our magnetometer, 
and we call it the {\it coherent probe} everywhere in the text. Such a
probe allows us to see the shot-noise limit (standard quantum limit (SQL)
of our magnetometer.


\begin{figure}[h]
	\begin{center}
		\includegraphics[width=1.0\columnwidth]{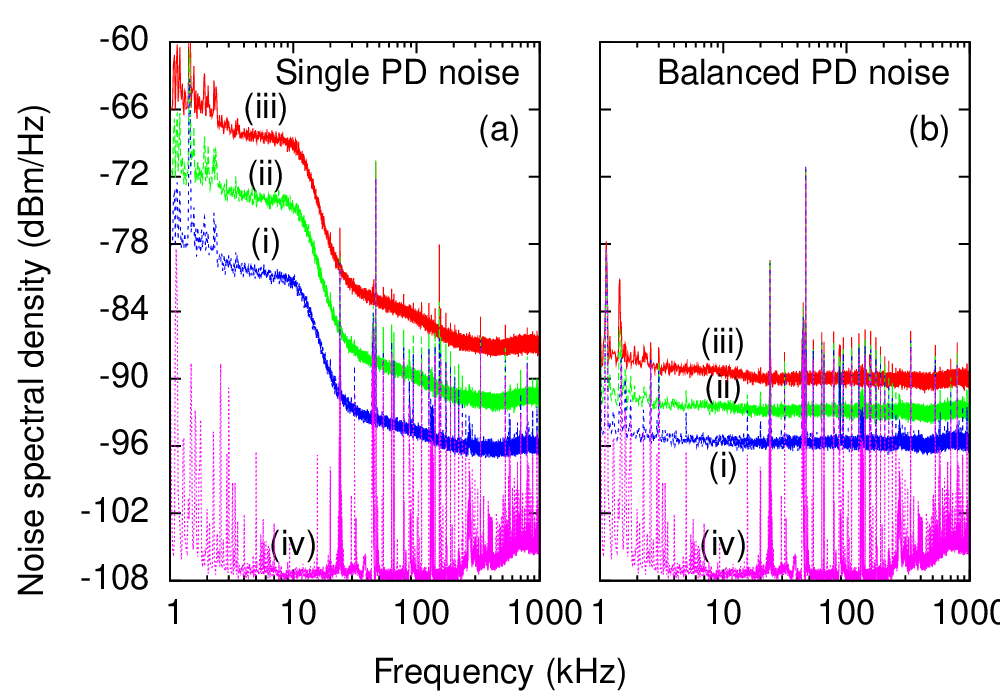}
	\end{center}
	\caption{(Color online) Comparison of the noise power spectral density of
	the laser residual
	intensity noise detected by a single photodiode (a) and balanced PD (b) for
	different laser intensities. Intensity of the laser doubles 
	between subsequent traces (i), (ii), (iii). The bottom trace (iv) corresponds
	to the
	dark noise of the detector.
	\label{fig:rin_shot_noise}
	}
\end{figure}

Unfortunately, our laser had a rather large  intensity noise and thus was not
shot-noise limited  along the x-polarization. We demonstrate  this by
inserting a
PBS into the squeezed field, bypassing the magnetometer cell, and directing
the laser  to detection  with one PD of the balanced setup blocked from
the light. In
this  configuration,  the  detector  is sensitive  to  the  amplitude  noise
quadrature of  the x-polarization  of the  probe field. As  can  be seen
in  Fig.~\ref{fig:rin_shot_noise}(a), the  noise  spectrum is  not flat
and
increases by  6~dBm/$\sqrt{\text{Hz}}$ at every subsequent  doubling of the
laser power: traces (i), (ii), and  (iii). In other words, the noise
spectral
density scales as the square of the  laser power, which is a signature
of  residual intensity noise (RIN).  However, 
our BPD detection  is shot
noise  limited  at most detection frequencies,
and  we  detect  noise  at the level of the
standard  quantum  limit
(SQL).  To prove  this,  we  open both  PD  of  the BPD  and  carefully
match  beam intensities  arriving to each.  As can  be seen 
in
Fig.~\ref{fig:rin_shot_noise}(b), the  spectral density now  scales
linearly
with the laser beam power,  i.e. it increases by 3~dBm/$\sqrt{\text{Hz}}$ at
each doubling of the laser beam power (see traces (i), (ii), and (iii)). The
noise  spectrum is  generally flat  with exceptions  of the  resonant noise
peaks  from  the electronics  (compare  to  trace (iv)  depicting  the
detector dark noise).  Comparing traces in Fig.~\ref{fig:rin_shot_noise}(a),
where one PD is blocked,  and Fig.~\ref{fig:rin_shot_noise}(b), where both PD
are open,  we see that we  can easily achieve about  25~dB RIN suppression.
Unfortunately, this is insufficient for truly shot noise limited
detection
at frequencies  lower than 200~kHz, indicated by the small  rise above the
SQL  level of  the noise  spectral density  at such  frequencies (see  also
Fig.~\ref{fig:lf_sqz}).  Therefore, while our squeezer offers noise 
suppression at detection frequencies as low as 100-200~Hz, we are only 
shot noise limited to start with in this experiment
at frequencies above 200~kHz due to laser noise.

\begin{figure}[h]
	\begin{center}
		\includegraphics[width=1.0\columnwidth]{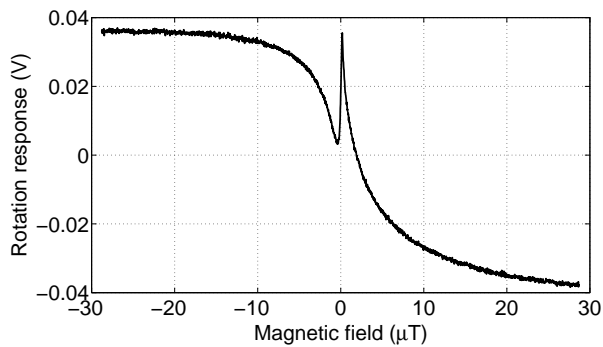}
	\end{center}
	\caption{Sample of the magnetometer response to the longitudinal magnetic field. The narrow feature at zero field is due to repeated coherent interactions of atoms with the light field.  Cell temperature is $40^{\circ}$~C, density is $6\times10^{10}$ atoms/cm$^3$, and probe power is 6~mW.
	}
	\label{fig:mag_response}
\end{figure}

\begin{figure}[h]
	\begin{center}
		\includegraphics[width=1.0\columnwidth]{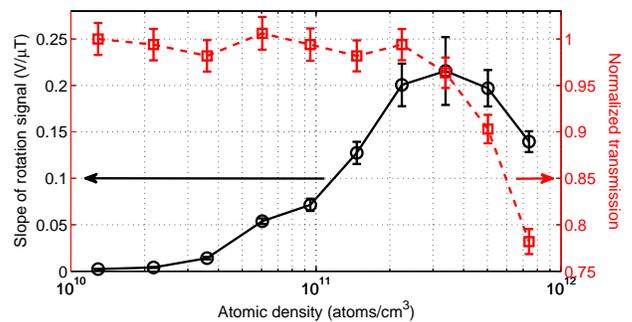}
	\end{center}
	\caption{Magnetometer response (solid) and probe transmission (dashed) vs atomic density.  Density uncertainties due to temperature fluctuations correspond to the size of the markers.  Laser power is 6 mW.  Cell temperatures range from 25-70$^{\circ}$~C in 5 degree increments.
	}
	\label{fig:mag_response_vs_n}
\end{figure}

When we  apply a longitudinal  magnetic field to  the magnetometer  cell, the
polarization of the probe field rotates  due to the NMOR effect and the
photodiodes
detect a signal proportional  to the angle  of rotation  (for small  angles) and
the incident intensity  of the light.  We fix the  intensity of light; thus the
BPD
signal is proportional only to the angle of rotation. A characteristic response
curve is depicted in  Fig.~\ref{fig:mag_response}. The broad S-like response
is governed  by the  Zeeman splitting  of the  ground-state magnetic  sublevels
and
their  decoherence time subject to power broadening 
(time of flight  of the atom, in  the  probe beam  is 
estimated to be around 3.3~$\mu$s, which
corresponds to a resonance width of 300~kHz, which in turn, governs
the relevant Zeeman splitting  to be  about 50~$\mu$T  for our  S-like resonance).  
The  narrow  resonance  at  zero  magnetic  field  is  due  to
velocity  changing  collisions  and  repeated  interaction  of  the  atoms
that  diffuse  away from  the  laser  beam and  then  return  back to  the
beam~\cite{zibrov'01ol, novikova05josab}. We attribute the asymmetric shape
to the presence of other hyperfine levels nearby that break symmetry.
 For such  atoms, the effective lifetime in
the beam  is significantly longer, resulting in a narrower spectral feature.  We
note
that if  we reduce the  power of the  probe beam below  1$\--$2~mW,  the  narrow  
resonance disappears, since the probe beam intensity drops below that required
to saturate the narrow resonance.
The  smallest detectable magnetic field
(i.e.  sensitivity)  of  the
magnetometer is  inversely proportional to the  slope of  this curve; the slope
is measured on the steepest part of the response curve on the left side of the
narrow peak.  This narrow
resonance thus increases  the response  of  the  magnetometer to very  small
magnetic fields, and so we maintain the intensity  of the probe light at
the level of several milliwatts. An easy way to boost the response of the
magnetometer is
to increase the number of interacting atoms in the magnetometer cell (i.e.
increase the cell temperature). The rotation signal slope (and thus
the magnetometer response) grows with increasing density
for small atomic densities (see Fig.~\ref{fig:mag_response_vs_n})
but then tends to saturate  since with increased atomic density the probe beam
is attenuated which leads to a weaker signal at the BPD~\cite{NovikovaMW02}. 
If the density is increased
even further, the  probe light will eventually be totally absorbed and 
no response will be detected.

\begin{figure}[h]
	\begin{center}
		\includegraphics[width=1.0\columnwidth]{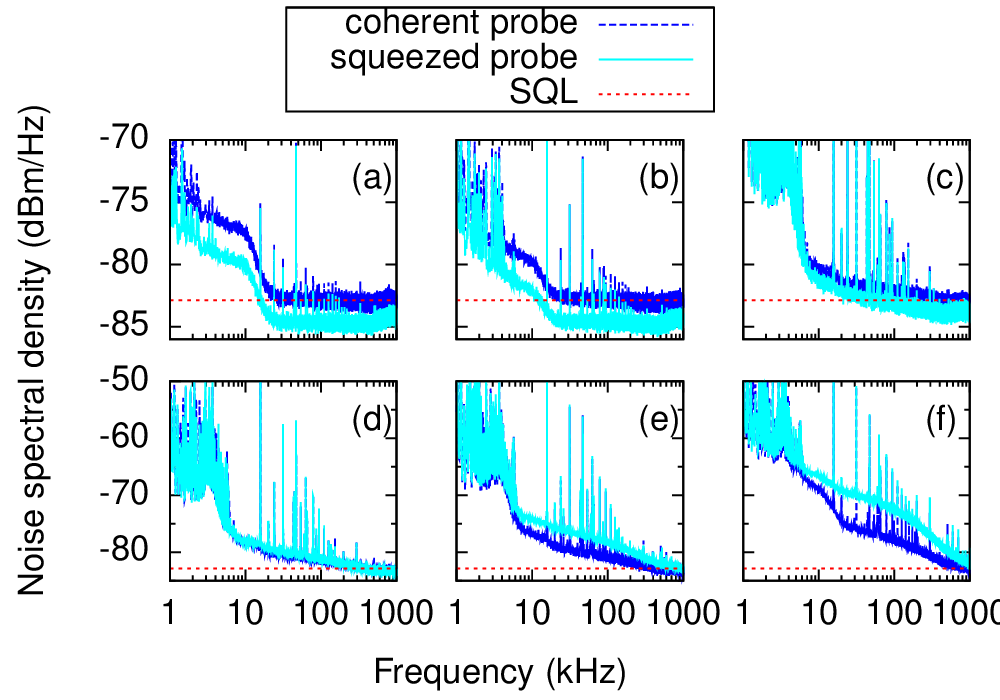}
	\end{center}
	\caption{(Color online) Magnetometer  quantum-noise-floor spectra  with
polarization-squeezed (light trace)  and
shot-noise-limited probe (dark trace) fields taken at different
temperatures/atomic
densities of the magnetometer. (a) 25$^\circ$C ($N=1.3\times10^{10}$cm$^{-3}$), (b) 35$^\circ$C ($N=3.6\times10^{10}$cm$^{-3}$), (c)
50$^\circ$C ($N=1.5\times10^{11}$cm$^{-3}$), (d)
55$^\circ$C ($N=2.2\times10^{11}$cm$^{-3}$), (e) 60$^\circ$C ($N=3.4\times10^{11}$cm$^{-3}$), (f) 70$^\circ$C ($N=7.4\times10^{11}$cm$^{-3}$).
Laser probe power is 6~mW.  
Spectrum analyzer resolution bandwidth is
28.6~Hz, the resulting trace is averaged over 300 traces.}
	\label{fig:mag_noise}
\end{figure}

The ultimate sensitivity is governed by the signal-to-noise ratio 
according to the equation $\delta B_z = (\partial \phi /\partial B_z)^{-1} \delta \phi$,
where  $\partial \phi /\partial B_z$ is the slope of rotation
and $\delta \phi$ is the noise level.  In our experiment, we use
the signal and noise of the voltage response of the oscilloscope, 
which is directly proportional to the angle of polarization rotation.
The noise level is set by the quantum
noise floor at frequencies higher than 200~kHz.
We compare the noise floors of our
magnetometer under two experimental conditions:  first, when we probe with
unsqueezed coherent light, which sets the shot-noise limit,
and second, when we use the polarization-squeezed light probe. We
conduct this comparison at different temperatures and atomic densities. The
results are depicted in Fig.~\ref{fig:mag_noise}.
During these measurements, we modulate the internal longitudinal magnetic field
at various
frequencies to ensure that noise floor of the magnetometer is unaffected by
the presence of alternating magnetic field. In 
Fig.~\ref{fig:mag_noise}, the noise measurements were taken
without magnetic field, but note the noise spike due to modulation set to
220~Hz for data in Fig.~\ref{fig:lf_sqz}.

At lower  atomic densities (cell temperatures),  when polarization coupling
does  not contribute  much to  the overall  noise budget,
we  see  broadband  noise  suppression  of  about  2~dB  from  hundreds  of
hertz  to several  megahertz, which  is independent  of atomic  temperature
and  follows the  input  squeezed  state noise  spectrum  (see for  example
Fig.~\ref{fig:lf_sqz},  obtained  with the  most  careful  balancing of  the
detector).  High  resonance-like  peaks  are  due  to  resonant  spikes  in
electronic dark noise  of the BPD and the electronic  noise of our solenoid
current  source. Note  that  in Fig.~\ref{fig:mag_noise},  one  can see  an
increase of  the noise  above the SQL level at  frequencies below  200~kHz and
especially below 10$\--$20kHz. This is due to residual intensity noise (RIN) of
our laser, discussed  above, making our detection not truly shot-noise limited at these frequencies, even  with the most careful balancing of the
light power at the PDs.

With increase of the atomic density in the magnetometer cell, we see that at
lower  frequencies  noise  grows  significantly above the SQL  level.  The
similarities  between the  RIN power  spectra   (Fig.~\ref{fig:rin_shot_noise})
and the magnetometer spectra (Figs.~\ref{fig:mag_noise}(d),(e) and
(f))  lead us  to  conclude  that this  contribution  of  the noise is from the
strong
x-polarization of the probe.  This contribution  is due  to the  dependence of 
the NMOR
effect on  probe power, thus  the RIN  in the x-polarization couples  into the
y-polarization noise that our BPD detects,
due to the presence of the atoms. To test this,
we block the y-polarized light with a PBS after the 
magnetometer and compare this noise floor to that of
the probe beam when it completely bypasses the atoms
in the magnetometer.  We find these noise levels are the
same, (adjusted for optical losses in the cell),
indicating that the increase in noise at high densities
is due to the x-polarized noise coupling into the 
y-polarized field.
However, we
note some interesting  dynamics: the squeezed probe shows a higher noise floor
compared  to the  coherent probe,  where  squeezing was  replaced with a normal
vacuum
state  in the y-polarization  (see  Figs.~\ref{fig:mag_noise}(d),(e) and  (f)).
We  conjecture  that this is due  to  the back action of atoms on the probe quantum
noise, since
we are unable to bring the noise level of the squeezed probe below the coherent probe level no
matter how we adjust the squeezing angle.

\begin{figure}[h]
	\begin{center}
		\includegraphics[width=1.0\columnwidth]{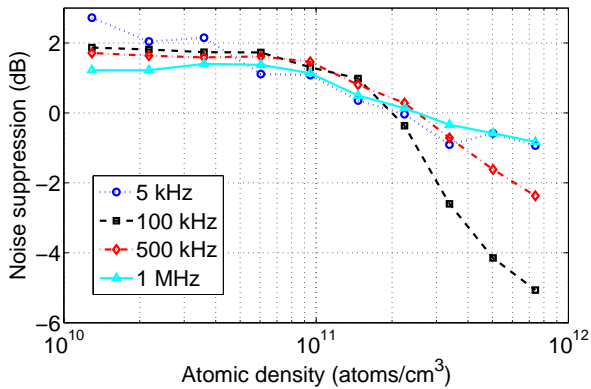}
	\end{center}
	\caption{
(Color online)
Noise suppression level vs atomic density normalized to shot-noise level
for several noise frequencies.  Positive values indicate noise suppression, 
negatives indicate noise amplification.  This level is found by averaging the
coherent probe noise level subtracted from the squeezed probe noise
level over 100 points (2~kHz) centered around the chosen noise frequency.  
The average uncertainty of $\pm 0.35$~dB is not included in the plot for 
clarity.  Laser probe power is 6~mW.
	}
	\label{fig:sq_vs_n}
\end{figure}

We   choose    several   noise    spectral   frequency    components   from
Fig.~\ref{fig:mag_noise}   to   better   illustrate   this   situation   in
Fig.~\ref{fig:sq_vs_n}. Here, 0~dB indicates the noise level 
seen using the coherent probe (unsqueezed
state).  
Note that at  lower atomic densities, the squeezing
clearly  improves the  magnetometer  noise  floor and  the noise  spectrum is 
nearly
independent of the detection frequency. At higher  densities, squeezing is
degraded due to absorption
by the atoms and so we expect less noise suppression.  
We also see that at the highest densities, 
due to the back action of atoms (as we discussed above),
the total noise is amplified rather than suppressed.
This effect shows that using squeezed light will
only improve the magnetometer sensitivity at certain
atomic densities and experimental conditions.

\begin{figure}[h]
	\begin{center}
		\includegraphics[width=1.0\columnwidth,angle=0]{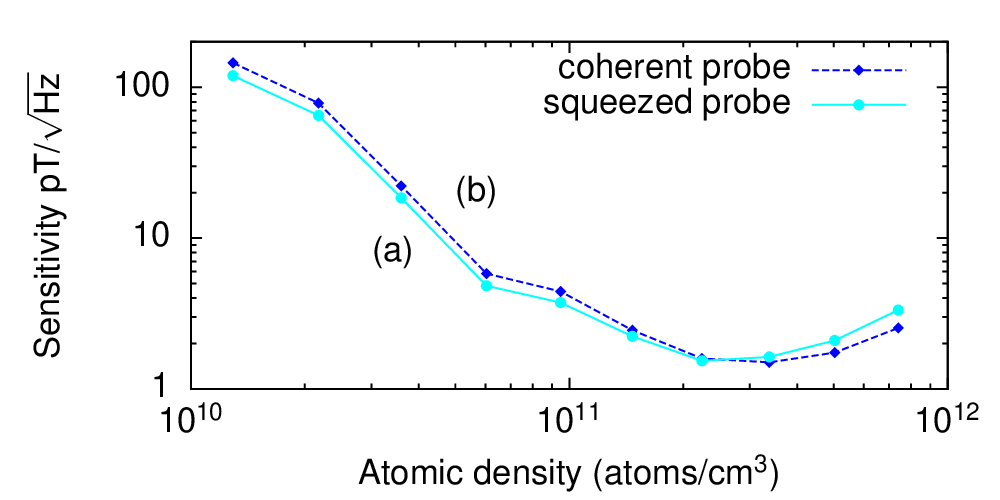}
	\end{center}
	\caption{
	(Color online)
	NMOR magnetometer sensitivity as a function of the atomic density
	with  polarization-squeezed (a) and  coherent (b) (shot-noise-limited)
	optical probes.  Errorbars are smaller than the size of the markers.  
	Laser probe power is 6 mW.  Detection frequency is 500~kHz.
	}
	\label{fig:mag_sens_vs_n}
\end{figure}

We  calculate   the  magnetometer  sensitivity  by   dividing the
noise  amplitude
densities (calculated  from the data presented  in figure~\ref{fig:mag_noise}) taken at 500 kHz, by the magnetometer 
response  shown  in  Fig.~\ref{fig:mag_response_vs_n}.
Due to absorption and the increased noise described above,
the NMOR magnetometer does not
benefit from polarization squeezing at all atomic densities and temperatures
as we show in Fig~\ref{fig:mag_sens_vs_n}.
However, benefits of the polarization-squeezed state probe are
clearly visible at lower atomic densities for the chosen detection frequency.
The magnetometer sensitivity can likewise be improved for any set of parameters
(detection frequency, atomic density, etc.) where noise suppression below
shot noise is observed.

We  demonstrated an  all-atomic  quantum  enhanced  NMOR   magnetometer  with
sensitivities down to close to
 1~pT/$\sqrt{\text{Hz}}$.  To the  best of  our
knowledge, this is  first demonstration of a squeezer at  795~nm capable of
noise  suppression below shot-noise levels at low  frequencies starting
from a few hundred hertz. This brings such a quantum-enhanced magnetometer into
the realm
of  practical applications  in  medicine and biology
where the characteristic magnetic signatures are  at sub-kilohertz frequencies.
We also  note that any DC  magnetic field  can  be up-converted to  the
detection band of this device if one spins the overall setup to generate a
modulation of
the magnetic field at the desired frequency. This may not be very practical for
an Earth-based setup,  but could be possible for a space-based setup,
where the overall rotation can be achieved at frequencies of hundreds of
hertz. So this enhancement method could potentially be applied to magnetometers 
used in astrophysics and space exploration programs. We also note that the increase in
noise below 200~kHz frequencies in our squeezer is not fundamental, and
can be improved with the use of a laser with less intensity noise and an improved
design of the BPD.
We would like to mention that our enhancement works
for any shot-noise-limited detection, and address a common argument against
squeezing that ``it is always possible to increase the SNR by
increasing the light power, making squeezing unnecessary.''  While this is
correct
{\em the injection of squeezing increases the SNR even further on top of the
power-boost improvement}.

\begin{acknowledgments}
We are thankful to Irina Novikova for useful discussions during the
manuscript preparation. Jonathan P. Dowling and Robinjeet Singh would like to
acknowledge support from the Intelligence Advanced Research Projects Activity
and the National Science Foundation.

\end{acknowledgments}

\end{document}